# Power, Preferment, and Patronage: Catholic Bishops, Social Networks, and the Affair(s) of Ex-Cardinal McCarrick

**Stephen Bullivant***
*Professor of Theology and the Sociology of Religion*
*St Mary's University, London, UK*

**Giovanni Radhitio Putra Sadewo**
*Research Fellow in Social Network Analysis*
*Swinburne University of Technology, Melbourne, Australia*

**\*Corresponding author:**
**Address:** St Mary's University, Waldegrave Road, Twickenham, TW1 4SX, UNITED KINGDOM
**Email:** stephen.bullivant@stmarys.ac.uk

## Abstract

Social Network Analysis (SNA) has shed powerful light on cultures where the influence of patronage, preferment, and reciprocal obligations are traditionally important. We argue here that episcopal appointments, culture, and governance within the Catholic Church are ideal topics for SNA interrogation. This paper presents prelminary findings, using original network data for the Catholic Bishops' Conference of England and Wales and the United States Conference of Catholic Bishops. These show how a *network-informed* approach may help with the urgent task of understanding the ecclesiastical cultures in which sexual abuse occurs, and/or is enabled, ignored, and covered up. Particular reference is made to Theodore McCarrick, the former DC Archbishop recently "dismissed from the clerical state". Commentators naturally use terms like "protégé", "clique", "network", and "kingmaker" when discussing both the McCarrick affair and church politics more generally: precisely such folk-descriptions of social and political life that SNA is designed to quantify and explain.

# Introduction

In recent decades, the collection of tools, theories, and methods commonly described as Social Network Analysis (hereafter "SNA") has – paralleling similar developments in both natural and computer sciences – shed powerful light on diverse areas of social, cultural, political, intellectual, and religious life. These include the epidemiology of contagious diseases (Keeling and Eames, 2005), the formation and spread of rival philosophical schools in medieval Europe (Collins, 1998:451-522), and collaborations among twentieth-century British classical composers (McAndrew and Everett, 2015). Undergirding much of this empirical work is an avowedly *relational* understanding both of individual subjects, and of the various, overlapping social worlds which they collectively inhabit, create, maintain, and/or change. Accordingly, if – as argued by one of relational sociology's leading proponents – "The most appropriate analytic unit for the scientific study of social life is the network of social relations and interactions between actors (both human and corporate)", then SNA has established itself as one of the principal means by which "we can identify *mechanisms* within interaction, relations and networks which help to explain and understand events in the social world" (Crossley, 2011:1, 4).

This paper has two main purposes. Firstly, we seek to show how certain methods and theoretical insights within SNA might illumine a subject of critical and – in light of events and revelations, both recent and long-ongoing – urgent significance within the sociology of religion. While appreciably slower to "catch on" in this field than in many others, there is now a reasonably substantial body of literature applying SNA within the sociological study of religion (Everton, 2018). To the best of our knowledge, none of it has focused on episcopal networks, whether Catholic or otherwise. In fact, there does not exist a huge amount of empirical social research on Catholic bishops, either as individuals or as collective actors – albeit with some notable exceptions. Closest in spirit to the approach adopted herein is Melissa Wilde's (2007) application of Social Movement Theory to episcopal politicking during the debates of Second Vatican Council. In addition, there has been a handful of psychological studies of US bishops based on mailed questionnaires (e.g., Sheehan and Kobler, 1977; Schroeder, 1978), though these are now forty years old. The American Jesuit priest and journalist Thomas Reese, who holds a PhD in Political Science, published detailed studies of several aspects of the Church's internal "power structure" (e.g., 1989; 1992). Most recently, researchers at the Center for Applied Research in the Apostolate in Washington, DC, have published a new monograph based on surveys and/or interviews with a majority of current "Ordinaries"[1] (Fichter et al., 2019).

Our second intention is to present initial findings from our own exploratory, "proof-of-concept" application of formal SNA to Catholic episcopal networks. This encompasses **i)** a pilot study, using data from the (relatively small) Catholic Bishops' Conference of England and Wales. We used this, primarily, to test out methodological decisions regarding the sampling, compiling, and coding of the dataset, and also to assist with hypothesis generation. And **ii)** various analyses, using network data for each member of the United States Conference of Catholic Bishops, as of July 2018, which – as we argue – lend considerable support to the hypothesis, widely shared among commentators on the Catholic abuse crisis, that there is a critical *network dimension* both to the problem and, perhaps, to its solution. 'The key now is to uncover the networks within the clergy and episcopate' (Dreher, 2018a). We conclude by suggested various possibile areas for future, deeper research.

**Background and rationale**

The motivation for this present study comes from three different sources. The first is the recognition that SNA has proven specially transformative in exploring networks in which the influence of patronage, preferment, and indebtedness are typically important. These could include anything from the mafia in Italy or the USA (Varese, 2013; Mastrobuoni and Patacchini, 2012), Chinese political elites (Keller, 2016), the academic job market (Kawa et al., 2019), or appointments to corporate boards of directors (Koenig and Gogel, 1981). *Prima facie*, episcopal appointments, culture, and (therefore) governance within the Catholic Church are ideal topics for serious SNA attention (see also Pogorelc, *forthcoming*[2]). Note, for example, the following *sui generis* features (cf. Reese, 1984; O'Callaghan, 2007: 119-52; Allen, 2016a):

i) The special nature of the bishop-clergy relationship (e.g., "Clerics are bound by a special obligation to show reverence and obedience to […] their own ordinary"; *Code of Canon Law*, 273), which in purely practical terms extends far beyond ordinary employer-employee dynamic within the secular world (outside of certain forms of bonded or indentured labour);

ii) the crucial role that the Ordinary plays in identifying, encouraging, and mentoring (including appointments to certain senior roles), potential future bishops;

iii) the critical role that a priest's own Ordinary plays in championing the 'candidacy' of a specific priest to be considered for episcopal office;

iv) the role that certain other, influential bishops have in supporting or undermining a particular candidate (especially the local Metropolitan, certain prelates known to "have the Nuncio's – or Pope's – ear", the previous Ordinaries of a given See, and/or members of certain national or Vatican committees (cf. *Code of Canon Law*, 377);

v) the fact that, aside from an "intense eight-day training program" held annually in Rome for newly appointed bishops, the primary way in which "bishopcraft" is learned is through informal apprenticeship under one's own previous bishop(s), especially through particular official roles within the diocesan administration, or – more formally – as an Ordinary's auxiliary bishop.

Whatever the *theological* justifications for, and desirability of, these features, from the perspective of SNA one might plausibly hypothesize a number of potentially negative properties emerging from this situation. For example, these might include the potential for ambitious clergy (or seminarians) to actively seek the favour and patronage of their own (and/or other influential) bishops, or indeed for bishops to use the hope – or even promise – of preferment as a means of incentivizing or rewarding loyalty. It could result in certain "types" (in terms of personality, class, ethnic background, theological vision, etc.) of priests being favoured and/or formed, in line with the type of their own bishop, and perhaps of a wider episcopal "mould" or "culture". This homophilizing tendency would then be intensified by the fact that "how to be a bishop" is learned, in very great measure, through a process of imitation and socialization. It might lead to the creation of identifiable "factions" or "cliques" of bishops, bound by mutual bonds of *preferment* and favour, who act – formally or informally – in concert, and who each support and promote each other's protégés. Furthermore, given all this, it might feasibly create shared senses of solidarity among particular groups of bishops, such that if "one falls, we all do". To give a hypothetical example: suppose that the bishops in a given province had all served as vicars general, chancellors, and/or auxiliary bishops for each other, and had in turn (even absent direct or formal collusion; cf. Bourdieu, [1984] 1988:84-9 on similar dynamics at work within academic appointments) returned the favour by supporting the promotion of each other's chancery favourites. Should one of the senior bishops in this group then be rumoured to have committed crimes while in office, it is not hard to imagine how others in the network might seek a "quiet" solution to the problem, to prevent either themselves or their patrons becoming implicated, even if by association, to varying degrees.

Secondly, this *a priori* fittedness of an SNA-informed approach to understanding episcopal governance structures receives considerable *a posteriori* support of various kinds. As noted above, Wilde (2007) demonstrates the value in adapting methods and perspectives commonly used to study political movements in order to shed light on the often-shadowy world of church politics. Furthermore, in reporting on and analysing such topics, well-informed journalists and academics naturally speak in terms of "networks", "factions", "cliques", "lobbies", "protégés", and "patrons" – that is to say, precisely the kind of folk-descriptions of social and political life which SNA seeks to quantify and interrogate. To give a single illustration here, note these excerpts from veteran Rome correspondent Robert Mickens, writing for the *National Catholic Reporter*:

> The current system the church uses to seek out and appoint candidates for episcopal service is far too often based on cronyism inherent in an old boys' network […] The apostolic nuncio plays a major role in drawing up the terna of (the top three) candidates for a particular episcopal post. [… The] roughly 30 cardinals and other ranking prelates from around the world who are members of the Congregation for Bishops […] discuss and vote on the candidates. […] However, well before this happens, bishops, in too many cases, have already begun "grooming" someone – perhaps a star seminarian or their priest-secretary – to be a future member of their very exclusive club, the episcopal college. […] Customarily, the ordinary of [a] large diocese has a fairly good chance of pinpointing the man or men he wants as an auxiliary bishop. And if he's well connected in Rome, especially with members of the pertinent congregation, this major hierarch can often help advance an auxiliary (or another bishop friend) to head his own diocese. (2016)

Finally, and more specifically, existing analysis of the scandals engulfing the Catholic Church in the United States (as in several other countries) highlights the role that precisely these kinds of network dynamics may have contributed, directly and indirectly, to both individual and institutional failures (and/or crimes) in adequately dealing with accusations of sexual abuse. Recognition of the need to focus on *organizational* cultures and contexts in both diagnosing and treating the "sexual abuse crisis" *in toto* is not wholly novel (Keenan, 2011). Yet it was thrust to the fore in 2018 primarily, though by no means exclusively, by the revelations surrounding Theodore McCarrick, formerly Cardinal-Archbishop of Washington, DC. While the full details, including allegations of grooming and sexual abuse of both boys and young men,[3] are beyond the scope of this paper, and have besides received very wide media coverage (see, in detail, Altieri 2020), a number of network-relevant aspects are worth highlighting.

**The McCarrick Case: A Relational Perspective**

First, McCarrick's predilection for identifying select groups of "especially favored" (Goodstein and Otterman, 2018) seminarians and young priests – i.e., "young men under his authority in the Church" (Dreher, 2018b) – whom he showered lavishly with alcohol, flattery, handwritten notes, gifts, meals, overnight stays at his personal beach house, and prophecies of great futures in the Church. These he referred to as his "nephews" and encouraged them to call him "Uncle Ted".

Second, the unwillingness of subordinates within his diocese(s) to refuse dubious requests, or to speak out in other ways. Note here the testimony of Fr Boniface Ramsey, who was a faculty member at Newark's archdiocesan seminary during McCarrick's 1986-2000 tenure as Archbishop, and whose repeated, unheeded attempts at whistleblowing have now come to light. Regarding McCarrick's widely-known practice of sharing a bed with his seminarians – he would deliberately invite more "nephews" to his Jersey Shore beach house than there were beds to accommodate – Ramsey rhetorically asks: "what member of the faculty would approach the archbishop to tell him that it just wasn't right?" Commenting on a "sense of resignation" among the seminary faculty, Ramsey recalls his own first attempt at raising the issue with his immediate superior: "The rector knew exactly what I was talking about and promised to do what he could to stop it, after admitting that *he felt strung between his loyalty to his archbishop* and his realization that what the archbishop was doing wasn't right" (2019; emphasis added).

Third, McCarrick's acknowledged status as the "the *kingmaker* for appointments in the Curia and the United States" (Viganó, 2018: 8) among American bishops. This claim has featured prominently in reports following the 2018 revelations, along with questions concerning the extent to which other high-placed US bishops and/or cardinals in his "network" might have benefited from his championing (e.g., Dougherty, 2018; Dreher, 2018c). Nor is this mere *post-facto* speculation. Well prior to the scandals erupting in June 2018, McCarrick's outsized pull on episcopal placings – over and above the Cardinal-Archbishop of Washington's traditional clout – was generally acknowledged. His direct role as the "architect" of specific appointments was also regularly reported in the media (Palmo, 2016; Allen, 2016b). Importantly, however, this kind of influence is not something unique to McCarrick: the existence of powerful "bishopmakers" follows naturally from the episcopal selection process as described above. Certain bishops have reputations for exerting various types of sway, whether official or unofficial, to have their own favorites and protégés elevated to the episcopacy.

Fourth, the fact that serious allegations about McCarrick were widely known, and even more widely rumoured, among the highest echelons of the hierarchy for decades, but were dismissed, ignored, or – in at least two cases – paid-off by his previous dioceses with five- or six-figure settlements. This retrospective "everybody knew" aspect is, of course, a common theme in the exposure of high-profile serial sex offenders (e.g., Jimmy Savile). A small number of (relatively junior) USCCB members have gone so far as to accuse their fellow bishops of, at best, negligence and, at worst, complicity and conspiracy. Thus Robert Barron, an auxiliary bishop in the Archdiocese of Los Angeles, writes:

> [I]t seems numerous bishops, archbishops, and cardinals, both in this country and in the Vatican, knew all about McCarrick's outrageous behavior and did nothing in response to it; or, rather worse, they continued to advance him up the ecclesiastical ladder, from auxiliary bishop, to bishop of a diocese, to archbishop, and finally to cardinal. Even after he resigned from his post in Washington, DC, […] McCarrick continued to be a roving ambassador for the Church and a kingmaker in the American hierarchy – again, while everyone knew about his disturbing and abusive tendencies. (2019: loc. 104; see also Lopes, 2018)

Understandably, questions have been raised as to *what* specific bishops close to him, including members of a so-called "McCarrick caucus" (Allen, 2016b) whose church careers he appears to have helped, knew and *when*, and what they did (not) *do* and *why* (not). Cardinal Donald Wuerl, who became Archbishop of Washington in 2006, resigned in October 2018, in part due to pressure from reports that he had known of accusations against his predecessor for over a decade without acting. Wuerl initially denied these reports before, after irrefutable evidence was produced, apologizing for having "forgotten" he had known (Guidos, 2019).

Fifth, the Catholic Church's male-only priesthood means that, while Catholic sexual abuse is *not* exclusively homosexual in nature, sexual activity (whether abusive or not) among bishops, priests, and seminarians *ipso facto* is. Furthermore, while celibacy is demanded of all Latin-rite priests[4] and seminarians, straight and gay, simply *being* same-sex attracted is, in and of itself, officially problematic (see Congregation for Catholic Education, 2005). From a network-perspective, this combination of factors has important relevance for understanding the McCarrick case. Not least, there is clear potential for *mutually compromised* networks of homosexually active (or once-active) priests, such as McCarrick appears to have cultivated among his "nephews". (By contrast, illicitly *heterosexually* active bishops and priests – there are no shortage of examples – can only be so with those *outside* of the priestly networks.) The existence of "homosexual subcultures" within US Catholic seminaries or diocesan power structures, while understandably a sensitive topic, is well-established in the academic literature, as too are the disproportionately high numbers of same sex-attracted seminarians and clergy in the first place (Greeley, 2004: 42-6; Cozzens, 2004: 124-39). In itself, that same-sex attracted seminarians and priests might form friendship and support groups with those who can empathize with their trials is wholly unsurprising. But, combined with an intensely homosocial environment, a culture of secrecy and shame combined with legitimate fear for one's vocation or ministry should one be "outed" (Martin, 2018), and a special bishop-priest/seminarian relationship which, irrespective of its other virtues and/or theological rationale, is certainly open to exploitation, then the risk of other McCarrick-esque cases is certainly a real one.

And sixth, details from subsequent episcopal scandals have shed further light on the wider culture in which McCarrick thrived for so long. Most intriguing here is the case of Bishop Michael Bransfield, who retired from the Diocese of Wheeling-Charleston, West Virginia, in late 2018. He too stands accused of sexually harassing and assaulting seminarians and young priests under his authority. Investigators also uncovered hundreds of cash "gifts" made from his personal account before being routinely reimbursed from diocesan funds, amounting to hundreds of thousands of dollars (Boorstein et al., 2019). Personal payments of four- or five-figure sums were regularly made to other bishops, especially those in influential positions in America and Rome – a practice which, as it transpires, is perfectly common. As one veteran Vaticanista puts it: "the impression one gets from bishops' public statements is that very few of them thought anything was strange about the money going around. It's just what high churchmen do, at least in the US" (Altieri, 2019). Other journalists have pointed out that McCarrick, too, was known for his largesse, and indeed he and Bransfield worked closely together on the Board of a major US-based Church fundraising charity, the Papal Foundation (O'Brien, 2018; Flynn, 2019). Several bishops and cardinals who received Bransfield's checks have since made clear that these gifts came with no strings or conditions attached, and thus were in no sense "bribes". Maybe so. But as generations of social scientists are all too aware, cultures in which reciprocal gift-giving is an embedded practice invariably tend to produce complex (and networked) relationships of trust, indebtedness, solidarity, obligation, and counter-obligation, even if the actors are not themselves fully conscious of them. That said, one presumes that McCarrick himself, having earned a PhD in Sociology from the Catholic University of America in 1963, might not be wholly unacquainted with the classic theories of Malinowski and Mauss.

**Social Network Analysis: An Introduction**

Since the precise techniques and jargon of SNA are not typically familiar to sociologists and other scholars of religion (Everton, 2018: loc. 549 n. 6), we feel it may be helpful here to provide a brief primer before proceeding.

A social network is defined as a collection of a finite set of actors (nodes) and the relation/ties between them (Robins, 2015). Actors (nodes) are defined as discrete individuals or groups and usually they have characteristics or attributes. The terms actors and nodes will be used interchangeably in this paper. As for ties, it seems that there is no formal definition in the literature except that they establish a linkage between a pair of actors (Borgatti et al., 2013).

**Social Selection and Social Influence Processes.** The two basic processes that underlie the formation of a social network are social selection and social influence (Robins et al., 2001). A social selection process is defined as a process in which actors structure their networks on the basis of member attributes. For instance, a bishop developed an alliance with other bishops due to personal characteristic(s) shared such as a common theological view. Social selection predicts that individuals select a certain position in the social network due to their personal characteristics, hence forming a relationship with other individuals that allow them to occupy that particular position. Conversely, social influence is a process in which actors gain/change certain attributes due to the network structure. Social influence may arise when individuals change others' behaviour or characteristic, or individuals imitate the behaviour or characteristic of others. An example of this would be if a bishop changes his theological view to conform to his colleagues. These two processes are usually intertwined; however, the focus of this study is only on social selection processes.

**Network Centrality.** Social selection processes would determine the nodes position in the network. Nodes in a central position tend to be regarded as having more prestige, influence, power, autonomy, and so on. Studies found that people in a central position tend to reap certain benefits such as having better grades (Bruun and Brewe, 2013), higher personal accomplishment (Shapiro et al., 2015), better organisational citizenship behaviour that leads to better well-being (Tsang et al., 2012), and so on. People in the central positions of the network might also determine the life and death of the network. In criminal networks, for example, it is sufficient to take out the key players to pacify the whole entire network (Sparrow, 1991).

There are several measures of centrality. In this paper, we would like to focus on degree centrality. Degree centrality refers to the number of edges (or ties) that a node (or vertex) has (Robins, 2015). The higher the number of edges/ties, the higher the degree centrality of the node/vertex. Nodes with high degree centrality would be highly visible and be considered as important in the network. Most of the information that runs through the network would pass by these nodes, giving them a clearer picture of what is happening in the network. Decision makers, leaders, social influencers tend to be in such positions. Based on this premise, bishops/cardinals that have a high degree of centrality in a network would have more power or influence to control the network and they will have more knowledge about the network. The high number of people that they have direct contact with would mean that they have the potential to diffuse information to the network quickly.

In a directed network, degree centrality is divided into indegree and outdegree. Indegree is the number of ties that a node received, while outdegree is the number of ties that a node sent. The interpretation for indegree and outdegree is quite different. In a network diagram of which bishops have served under whom, a high indegree would mean that a bishop had a lot of others serving under him, while high outdegree would mean that this bishop had served under many bishops. Our focus is directed into indegree centrality as we theorize that when a certain bishop has high indegree, he would have more power over the network as he has more people in the network that was "moulded" by him. There are still many other measures of centrality such as betweenness, beta centrality/bonacich, closeness, and k-reach centrality, but those are beyond the scope of this paper.

## Pilot study: England and Wales

Before creating the USCCB dataset, we felt it important to do a test run based upon a much smaller dataset. This would allow a number of basic methodological choices regarding the scope (people, timeframe) and coding (what types of "tie", and how differentiated) of the underlying adjacency matrix to be honed effectively: repeat "do-overs" are not so easy with a manually compiled 422 x 422 matrix, as our US dataset turned out to be. The Catholic

Bishops' Conference of England and Wales was selected for this sandboxing stage, due both to its size (with 27 dioceses and equivalents,[5] compared to the US's 197), and to one of us being familiar with its personnel and reported dynamics.

Naturally, there are many ways in which bishops might be socially "tied" to each other: e.g., attending the same seminary or university, being involved with certain apostolates, or simply through getting to know each other at various (formal or informal) get-togethers. This is particularly true of the English and Welsh Church, which is a small world both geographically and otherwise. For reasons outlined above, however, i) the special, *sui generis* relationship between an Ordinary and his literal subordinates, and ii) the significant influence of an individual's current and former Ordinaries over whether he becomes a bishop himself, and if so, over his subsequent episcopal "career", appear to be of overriding importance. This Ordinary-subordinate "tie" is not a uniform one: it is reasonable to hypothesize differing "strengths" to the relationship based on different roles that a given priest/bishop has served in under a given bishop.

Given these assumptions, the basic sample was restricted to all living bishops who either are, or have been (i.e., *emeriti*, many of whom remain active in pastoral and administrative work), members of the Catholic Bishops' Conference of England and Wales. This excluded any papal diplomats *from* England and Wales who have been ordained as (arch)bishops, but who are not and never have been CBCEW members. It included, however, one bishop (Roche) now assigned to a fulltime role in the Roman Curia, but who was previously the Bishop of Leeds. Data was originally collected in mid-October 2018, then slightly updated in October 2019 for publication in order to reflect a number of new retirements and appointments in the intervening year. Ties were defined according to which bishops the members of this original sample had *served under* (including any now deceased, though these are not included in the below network map), coded for three different "weights" of seniority/trustedness of role:

1) whether A has served directly under Ordinary B in some capacity (i.e., as a priest of B's diocese, or – in the case of Bishopric of the Forces – as a priest on loan to the Ordinary);

2) whether A has served in one of a small set of especially high-trust diocesan positions (Vicar General, Episcopal Vicar, Chancellor, seminary Rector, Private Secretary within diocese) under B, *or* as General Secretary of the Bishops' Conference under B as Cardinal-President.

3) whether A has served as an Auxiliary or Coadjutor Bishop under B.[6]

Where A has served B at one or more different levels – e.g., a stint as an Episcopal Vicar (2), following on from many years as a regular parish priest (1) – the tie was recorded in accordance with the highest level (so 2). Given the precedence accorded to the Ordinary-subordinate relationship, cases where, e.g., C was a religious priest serving in a parish in B's diocese, but not *under* B's canonical authority (that is, C's Superior remained his own Abbot or Provincial), were not counted as a tie (though serving under B at levels 2 or 3 would be). These data were collected and inputted manually into Excel, creating a 80 x 80 adjacency matrix (including those deceased). One virtue of this coding methodology is that, in principle, these ties and their weights are based upon objective, publicly available data: whether or not B has served under A, and when and in what formal capacities, are facts of the kind that would feature in any "ecclesiastical CV". A variety of web-based sources were used, including biographies given on official CBCEW and diocesan websites, individual bishops' Wikipedia pages, and (for dates of ordination and episcopal appointments) http://www.catholic-hierarchy.org/.

Figure 1 shows the resulting network map of all living CBCEW members, active and retired (names in parentheses), constructed using the NodeXL software package (Smith et al. 2010). The graph was drawn using the Harel-Koren Fast Multiscale algorithm (Harel and Koren 2002), then lightly amended for ease of presentation. Edges are directed, with "A →  B" signifying "A served under B", and are assigned a tone/width corresponding to the three weights outlined above (with 1 being the lightest/thinnest, and 3 the darkest/heaviest). Nodes are sized according to their indegree (i.e., the total number of other bishops/nodes, irrespective of edge weights, who have served under them). Of 58 nodes, 40 form part of a single constellation, connected by 49 edges; a further seven are one of three much smaller constellation, of one or two edges each; 11 are isolates. The main constellation has a diameter of 9, with a mean distance between nodes of 3.8.

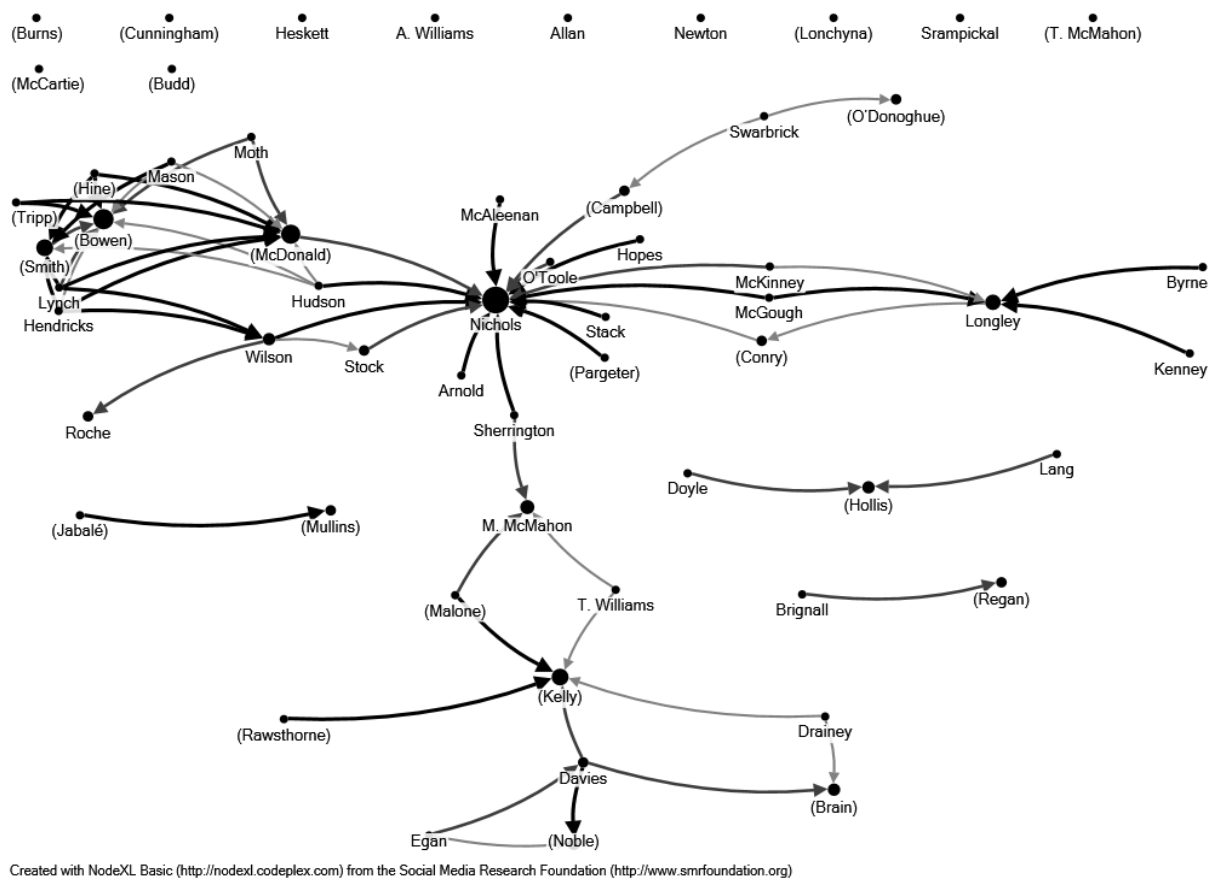

**Fig. 1. Network map of living members of the Catholic Bishops' Conference of England and Wales, both active and retired (October 2019)**

The purpose of this pilot study was simply to test out various sampling and coding strategies, and to see whether the application of SNA to a Bishops' Conference could produce a meaningful network map. This it certainly appears to have done. The influence of certain Metropolitan Archbishops, due both to their special role in the nomination process, and their typically having several auxiliary bishops under them, is clear: all 7 nodes with an indegree of 3 or more are, or were, Archbishops of Westminster (Nichols), Birmingham (Nichols, Longley), Southwark (Bowen, McDonald, Smith), Liverpool (Kelly, M. McMahon), and/or Cardiff (Smith). Furthermore, the graph clearly shows the overwhelming influence and centrality of Cardinal Nichols, with a quarter of all bishops (15 out of 58), and a third of all non-retired bishops (11 out of 34), having served under him, in most cases as auxiliaries. Nichols' centrality is, moreover, probably greater than that of his recent predecessors, having led another large archdiocese for nine years prior to his translation to Westminster (which neither Hume, 1976-99, nor Murphy-O'Connor, 2000-9, had).

The fact that the graph conforms, in these and other basic respects, to what any observer of English and Welsh Catholicism "could have told you anyway" is a good sign: a network map seeming to show, say, Nichols as a marginal figure within the CBCEW, or one of the long-retired auxiliaries as its kingpin, would be (correctly) suspected of having fatal methodological flaws. That said, the above graph does *not* simply represent a wildly convoluted and laborious method of arriving at obvious facts. It also, arguably, sheds light on the more subtle dynamics of CBCEW politics. Two examples may be given here. Firstly, it was reported that, prior to the 2016 EU referendum, a Nichols-backed attempt to put out an official CBCEW statement in favour of Remain (i.e., anti-Brexit) was thwarted by a minority of bishops (Thompson 2016). If one knew nothing more about the Bishops' Conference than the information presented in Figure 1, then a betting person might reasonably suppose this rebellion to have been fomented within the relatively dense "Southwark cluster" towards the left of the graph, most of whose members are at two removes from Nichols' direct circle. This was in fact the case, with Peter Smith, then Southwark's Archbishop, reportedly leading the charge. Secondly, there are two bishops whose occasional public statements have been widely perceived as being out of step with the CBCEW consensus on certain "hot button" issues (e.g., on the reception of communion by the divorced-and-remarried, or by politicians supporting abortion or same-sex marriage). Devotees of SNA will not be surprised to learn that these two, Egan and Davies, are a) themselves closely tied, with both having served under Noble, and Egan having been Davies' Vicar General; and b) out on a limb, in both senses, in terms of the wider CBCEW network (Egan, at 6 removes, is the joint-furthest vertex from Nichols in the main constellation).

## United States Conference of Catholic Bishops

### Methodology

In essentials, data collection and input followed the procedure laid down in the pilot study, with the exception that deceased bishops were automatically excluded from the sample. In total, there were 424 bishops in the US as of July 2018 (i.e., when the McCarrick scandal broke). We also collected data on whether they are cardinals, archbishops, bishops, auxiliary bishops, and whether they have retired or not.

As the first step in analysing the data, we undertook several visualisations of the network. We used two types of community detection algorithms, K-Core and VOS, using the software package Pajek (Batagelj & Mrvar, 2011). K-Core creates subnetworks of a given network where each node has at least $k$ neighbours in the same core. We experimented with different values of $k$ to find an optimal result and ended up with $k = 3$. VOS is a community detection algorithm that determines the clustering ($C$) of a network by maximizing its modularity. Below is the formula:

$$Q(C) = \sum_{C \in \mathcal{C}} \left(\frac{1(C)}{m} - \left(\frac{d(C)}{2m}\right)^2\right)$$

1(*C*) is the number of ties between nodes belonging to cluster *C*, and *d*(*C*) is the sum of the degrees of nodes from C.

The second step of the analysis was calculating the degree centrality of each individual bishop. Since it is a directed network, we calculated indegree. This calculation was done using UCINET (Borgatti et al., 2013). We then listed 10 bishops with the highest indegree (that is, the highest number of other bishops in the networks who have served under them).

## Results

Out of 424 bishops, there were 256 active bishops and 168 emeritus bishops. The mean age of the bishops is 71.12 years ($SD = 11.69$), with a mean time they were ordained as bishops of 17.61 years ($SD = 13.01$) The complete breakdown of the rank is provided in Table 1.

| Rank | Number | Percentage |
| --- | --- | --- |
| Cardinal-Archbishop | 6 | 1.4% |
| Cardinal-Archbishop Emeritus | 9 | 2.1% |
| Archbishop | 28 | 6.6% |
| Archbishop Emeritus | 19 | 4.5% |
| Bishop | 143 | 33.7% |
| Bishop Emeritus | 92 | 21.7% |
| Auxiliary Bishop | 78 | 18.4% |
| Auxiliary Bishop Emeritus | 48 | 11.3% |
| Coadjutor Bishop | 1 | 0.2% |

**Table 1. Ranks of bishop within USCCB sample, active and retired (July 2018)**

The next step of the analysis is creating visualisations of the network. These visualisations were produced using Pajek software. The first visualisation was produced using K-Core (fig. 2) and the second was using VOS (fig. 3). In both cases, active bishops are indicated by round nodes, and retired bishops by square ones, with all nodes sized according to their indegree (weighted, according to edge weight). Edges are colour-coded, according to the three strengths of Ordinary-subordinate relationships outlined above: with black for "1" (normal priest), blue for "2" (higher-trust/senior priestly roles), and red for "3" (auxiliary or coadjutor bishop).

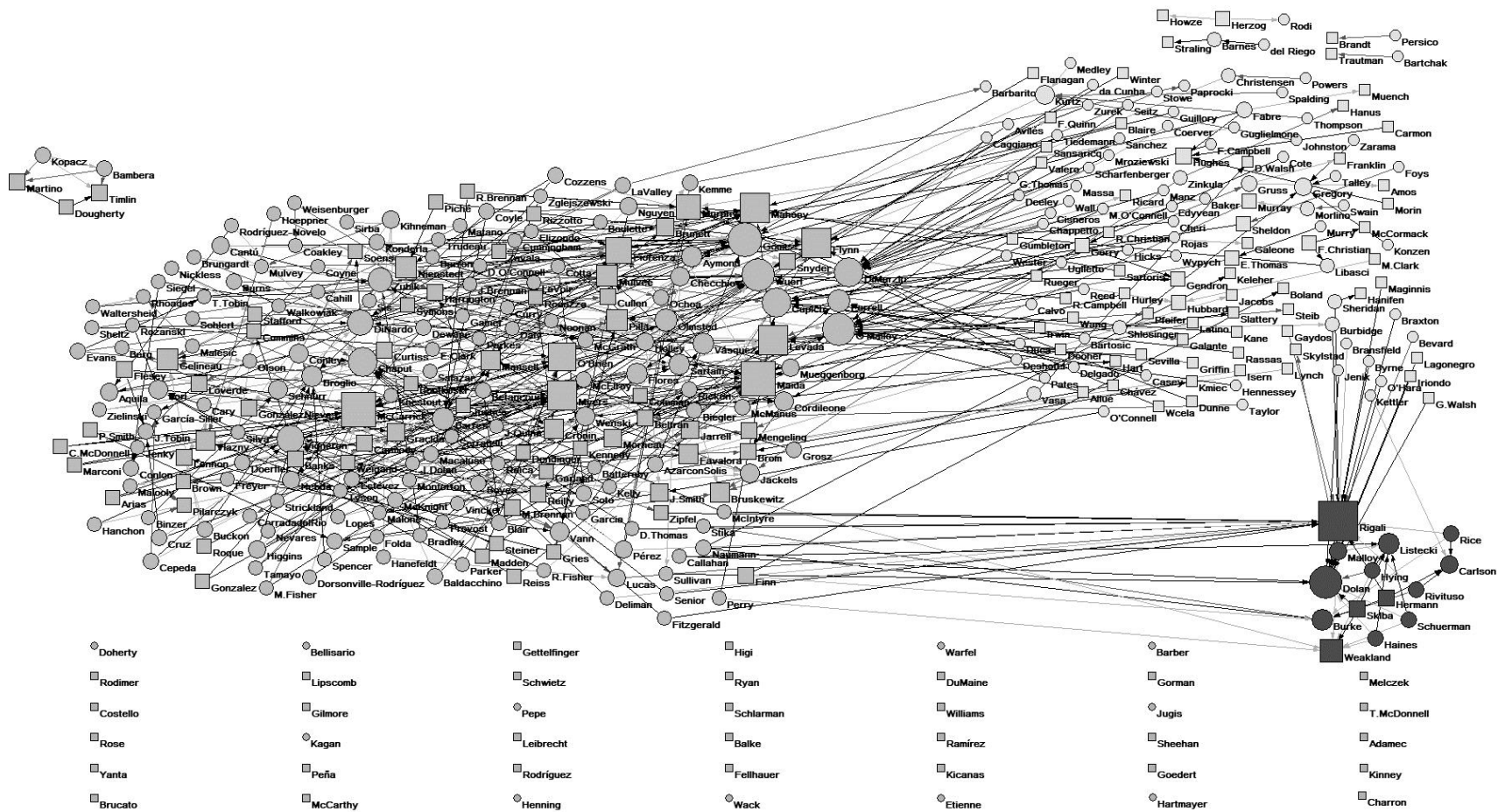

**Fig. 2. Visualisation of US bishops' network using K-Core (*k* = 3) (July 2018)**

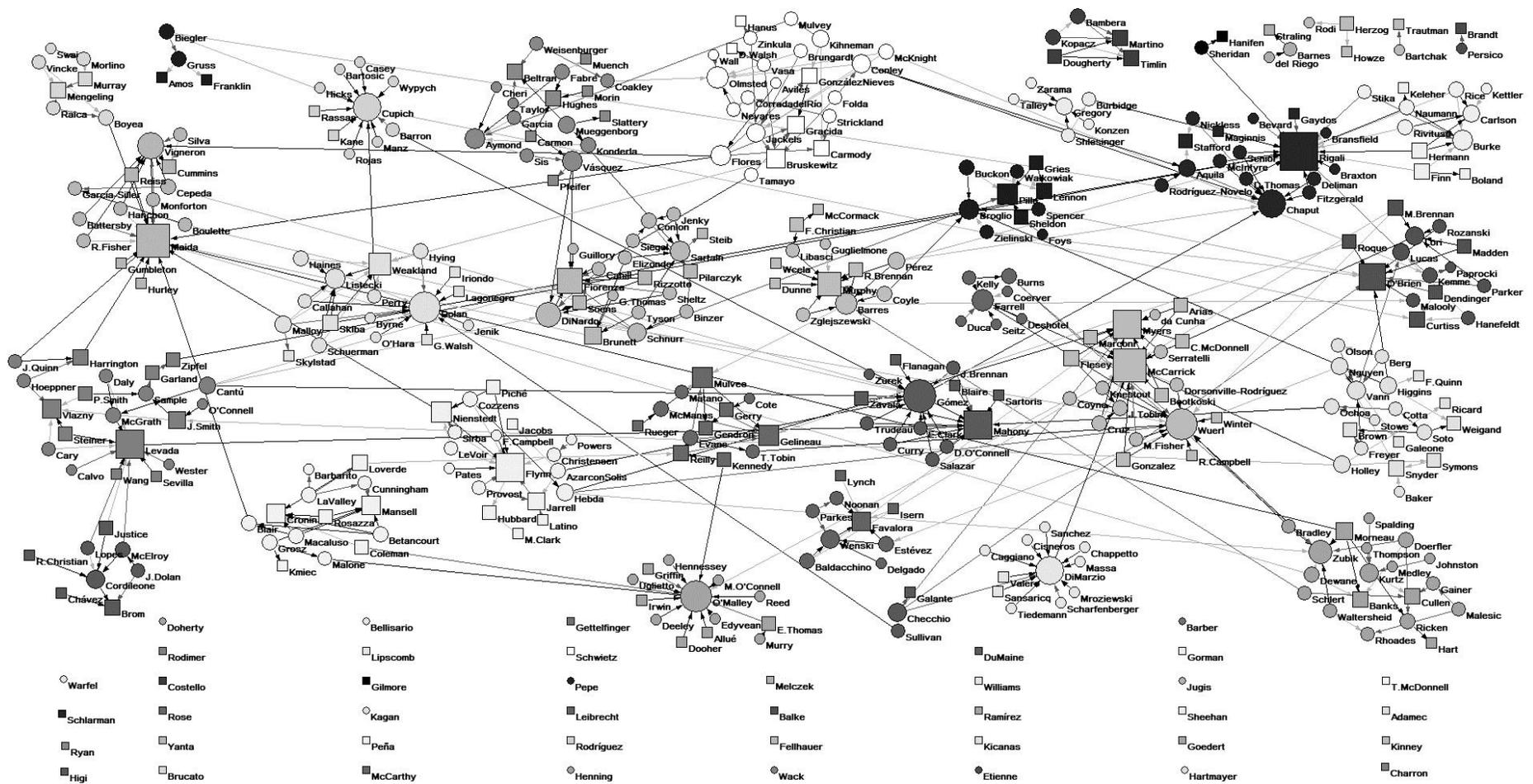

**Fig. 3. Visualisation of US bishops' network using VOS algorithm (July 2018)**

As is clear from the visualisations, there were some very large nodes (i.e., bishops who had high indegree). However, relying on sight alone, it would be difficult to determine accurately which bishops were in the center of the network (the key players). To obtain the information, we calculate the indegree of the bishops. The bishops have an average of 1.25 ($SD$ = 2.84) indegree. The top 10 bishops with the highest indegree are presented in Table 2.

| Name | Rank | (Last) Diocese | Indegree |
|---|---|---|---|
| Rigali | Cardinal-Archbishop Emeritus | Philadelphia | 22 |
| McCarrick | Cardinal-Archbishop Emeritus | Washington | 17 |
| Maida | Cardinal-Archbishop Emeritus | Detroit | 17 |
| O'Malley | Cardinal-Archbishop | Boston | 15 |
| Gómez | Archbishop | Los Angeles | 14 |
| Wuerl | Cardinal-Archbishop | Washington | 14 |
| Dolan | Cardinal-Archbishop | New York | 13 |
| Chaput | Archbishop | Philadelphia | 12 |
| Mahony | Cardinal-Archbishop Emeritus | Los Angeles | 12 |
| Cupich | Cardinal-Archbishop | Chicago | 11 |

**Table 2. US bishops with highest (weighted) indegree (July 2018)**

Given the nature of episcopal appointments, it is no surprise that nodes with highest indegree are (or have been) Ordinaries of large dioceses, which typically have a significant number of auxiliaries, and in many cases have been Ordinaries for a long period: the top three are all *emeriti* of one or − in the cases of McCarrick (Newark and DC) and Rigali (St Louis and Philadelphia) − two major Archdioceses. Note too the presence of three voting members of the Vatican's Congregation for Bishops, past (Rigali) and present (Cupich and − resignation from DC notwithstanding − Wuerl). Rigali's pre-eminence is also noteworthy, given his decades-long career within the Roman curia *prior* to his appointment as Archbishop of St Louis in 1994, including five years as second-in-command of the Congregation for Bishops.[7]

Finally, we extracted an ego-network for McCarrick, based on two degrees of separation (in either direction). Figure 4 presents a visualisation of this network, created with UCINET. So defined, McCarrick's "personal community" (Chua et al, 2011) includes a total of 43 bishops, including several who are significant nodes in their own right (e.g., Farrell,

DiMarzio, Wuerl, Holley, Myers, J. Tobin). Note also the existence of certain alter-alter ties, that is, direct ties *between* non-McCarrick nodes.

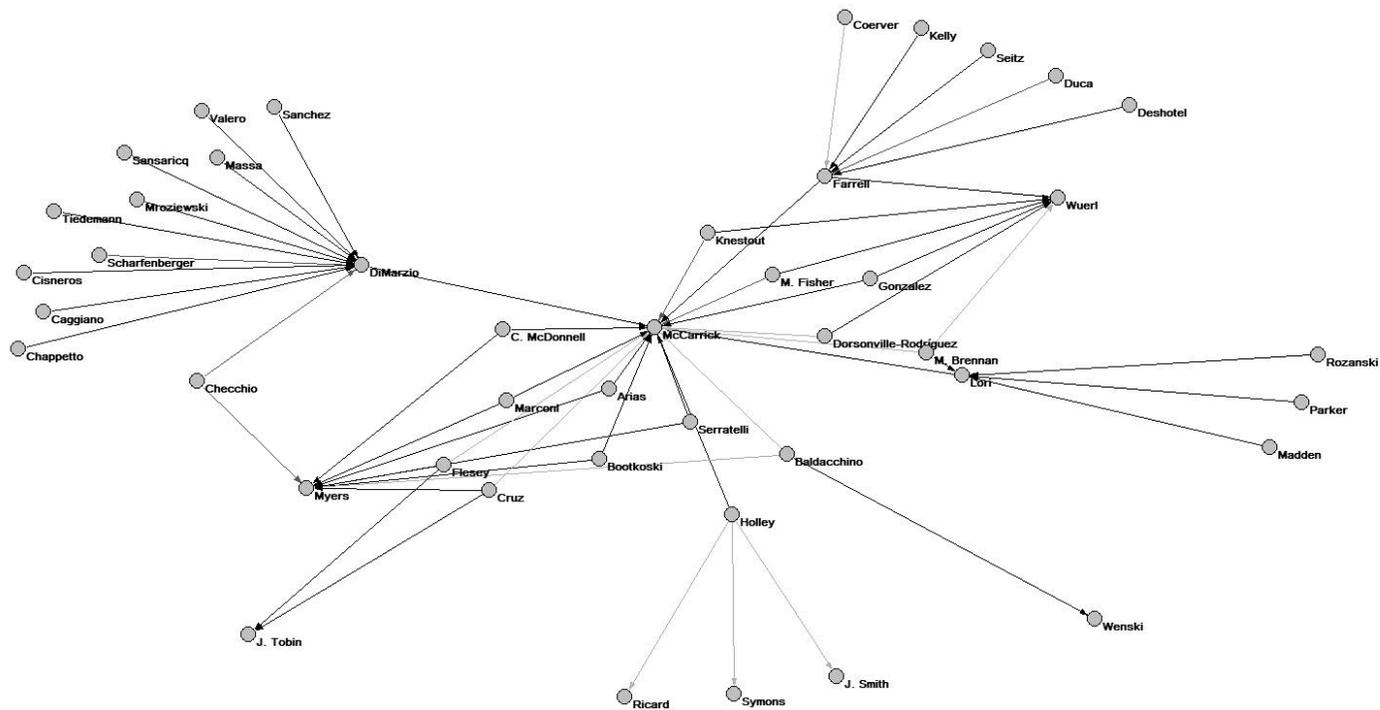

**Fig. 4. Ego-network for McCarrick, showing two degrees of separation (July 2018)**

**Discussion**

These exploratory studies offer an original contribution to several live issues regarding episcopal structures and governance. At the most basic level, our network maps support the view that it is meaningful to talk in terms of there being defined "cliques" of bishops. While previous discussions of this phenomenon have tended to be limited to considering the direct relationships between a single bishop and his various "proteges" (e.g., Dreher, 2018a), our network visualizations reveal more complex kinds of clustering. That is to say, whether bishops B, C, and D have both served under A is not the only factor to take into account. Important too is whether B, C, or D have served under each other, or under a bishop Z of whom bishop A was himself once a protégé (and so on). As is clear from fig. 3, the US bishops do indeed tend to cluster into various more-or-less dense "flocks".

It is important to state that network proximity to a McCarrick (or whomever) is not itself any indication of another bishop's knowledge of, complicity in, or imitation of a McCarrick's misdeeds. And indeed, we know from other cases that sexual offenders are often adept at hiding their crimes, including from those closest to them. Nevertheless, given the nature of both the clerical Ordinary-subordinate relationship, the outsized influence that a priest's own (former) Ordinary/ies has/have in advancing his own ecclesiastical career, and the great extent to which "bishopcraft" is passed on through a form of apprenticeship, then it would be naive in the extreme to ignore this "network dimension" entirely. We wish to highlight two significant issues where we think this kind of network-thinking might make a direct, positive contribution.

The first relates to several inquiries into "historical"[8] sexual abuse and its handling by church authorities, not least those commissioned by multiple states' Attorneys General. Apparently without exception, these reveal a "standard response to reports of abuse by church leadership" (Missouri Attorney General, 2019: 2) practised with remarkable consistency across different dioceses. As one scholar has recently summarized it:

> A more-or-less standard pattern emerged: credible allegations against an abusive priest being kept quiet, with assurances made to the victims and their families; the priest in question being quietly reassigned, perhaps after a period of 'successful' counselling at one of a small group of Church-run treatment centres specializing in precisely this, or sent to another diocese with a glowing letter of recommendation; no thought whatsoever being given to this new set of young people being put into very serious harm's way; and this process being repeated, multiple times, for years if not decades. (Bullivant, 2019: 225)

As the Grand Jury tasked with investigating six dioceses in Pennsylvania arrestingly puts it: "Special agents testified before us that they had identified a series of practices that regularly appeared, in various configurations, in the diocesan files they had analysed. It's like *a playbook for concealing the truth*" (Pennsylvania Attorney General, 2018: 3; emphasis added).

This begs the question of precisely *how* this unwritten, secret "playbook" came both to be and to spread. There is no evidence of any explicit conspiracy between Catholic dioceses to create a set of norms or procedures to be followed in such circumstances, and yet it also seems a stretch simply to suppose that almost exactly the same "solutions" arose complete independently, by spontaneous generation, in each chancery. Much more plausible, we contend, is to view this metaphorical "playbook" as a set of routinized practices and norms, or *habitus* (see Bourdieu, [1972] 2013; Archer, 2010), emerging and diffusing "organically" within and through ecclesial networks. SNA studies have explored the ways in which the creation and spread of social norms, of various kinds, are affected by differences in network structure (e.g., Sen and Sen, 2009). Feasibly, the combination of i) socially learned

"bishopcraft", with senior chancery roles serving as a quasi-apprenticeship for future bishops; ii) the existence of both strong, densely clustered groups of bishops in particular regions; and iii) weaker, though still significant, links *between* these local groups, due to rarer cross-country promotions of bishops (cf. Granovetter, 1973); provides the ideal conditions for this unwritten "playbook" to become informally codified. Also amenable to serious SNA explanation, at least in principle, is the "underground railroad" shunting problem priests between dioceses: a rigorous mapping of which bishops loaned abusive priests to which other bishops could be a fascinating study in its own right.

The second issue we wish to highlight is the scope for very serious *conflicts of interest*, especially among bishops with close ties to each other. The way episcopal appointments currently happen, it is frequently the case that bishops in a certain region are relatively densely clustered. This follows from several features (*not* bugs) of the system, not least the tradition of auxiliary bishops being created locally (with the Ordinary having a fairly free hand in nominating candidates),[9] and the role that a province's Ordinaries have in recommending candidates to nearby vacant sees. Undoubtedly, this system brings many benefits: for example, it is far more efficient than an unwieldy national system, and increases the chances of bishops actually knowing the people they are putting forward. One of several side-effects, however, is that if complaints are made against the former bishop of a diocese, then there is a strong likelihood of the *current* bishop being quite closely networked with him: even if neither has previously served under the other, the odds are good that they have mutual ties with other bishops who have. And indeed, this is precisely what happened with McCarrick. When the (arch)dioceses of Metuchen, Trenton, and Newark all made large cash settlements to McCarrick accusers in the mid-2000s, Metuchen's Bootkoski and Trenton's Smith (now deceased) had both been appointed by McCarrick as his auxiliaries. Meanwhile, Newark's Myers, while not directly tied to McCarrick, *is* tied to eight other bishops who served under McCarrick, including Bootkoski (see fig. 4). Likewise Wuerl, whose inaction when Archbishop of Washington was noted earlier, though also with no direct tie to McCarrick in fig. 4, nonetheless shares six former subordinates with him. Since new Ordinaries obviously "inherit" the auxiliaries and other clergy of their predecessors, these kinds of overlapping connections are impossible to avoid fully. It is not hard to think up scenarios in which bishops X and Y "sharing" subordinates , C, and D, even in the absence of a direct tie between X and Y, might create conflicted loyalties between the two. After all, if Y is found to have been engaged in serious wrongdoing, then questions will naturally be asked as to what knowledge or involvement B, C, or D might have had. Since B, C, and D may now be X's own closest collaborators, friends, and protégés, the temptations of a quiet, out of court, and/or (to use a favoured euphemism) "pastoral" solution to the problem may be irresistible. More invidious still, consider that B, C, and D may now have intimate knowledge of any "dirt" on X, which, if X goes public with accusations against Y, might itself be leaked.

This recognition has significant implications regarding new canonical norms issued by Pope Francis in 2019, specifically to address the question of how serious complaints against bishops are properly to be dealt with and investigated. One noteworthy feature is that primary jurisdiction is granted to a province's Metropolitan Archbishop to investigate allegations against bishops, past or present, in that province. (If the Metropolitan is himself the subject of the allegations, then this role would normally devolve to the "most senior suffragan Bishop by promotion, to whom, if such is the case, the following provisions regarding the Metropolitan apply"; *Vos Estis*, art. 8 §2). These canonical norms owe much of their genesis to the US crisis. Indeed, the fundamentals of this so-called "Metropolitan model" were widely reported to have been drafted by Wuerl and Cupich in the run-up to the USCCB's own plenary meeting held in November 2018, something they themselves denied (White 2018).

> The new norms are, explicitly, alive to the potential for conflicts of interest:
>
> The Metropolitan is required to act impartially and free of conflicts of interest. If he considers himself to be in a conflict of interest or is unable to maintain the necessary impartiality to guarantee the integrity of the investigation, he is obliged to recuse himself and report the circumstance to the competent Dicastery. (*Vos Estis*, art. 12 §6; see also 13 §3)

The practical difficulty, however, is this: given the structure of episcopal networks, there would seem to be a very low probability of any Metropolitan being "free of conflicts of interest" if asked to investigate other bishops, past or present, of his own diocese or province. Even where no direct ties exist, there will frequently be other close ties between mutual subordinates or superiors, protégés or mentors. And this is quite apart from other proprietary concerns that a Metropolitan might have about the moral, financial, and reputational liabilities of his own diocese and its personnel. Even apart from the kinds of Ordinary-subordinate ties which we have particularly privileged in this paper, given other kinds of ties between US bishops, one wonders how any senior bishop can ever be, and/or be seen to be, truly "free of conflicts of interest" when investigating "brother bishops". Indeed, this much is clear from the very first "Metropolitan model" investigation carried out in the US (*before* the model was enshrined in canon law): the above-discussed investigation of Bransfield by William Lori, Archbishop of Baltimore. As it turned out, Lori was himself one of the high-ranking prelates who received cash gifts from Bransfield: $7500 in total, including $5000 to mark his *becoming* Bransfield's Metropolitan. The original publication of Lori's report redacted the names of those who had received Bransfield's gifts, including – of course – Lori's own (Ferrone, 2019; Pogorelc, *forthcoming*). Furthermore, our own network data show that while Lori and Bransfield are not closely tied, Lori served, if briefly, as an auxiliary of McCarrick's in Washington. McCarrick too received cash from Bransfield. In what other area of life would oneself and one's close associates having received thousands of dollars in "gifts" from the person whose case one is investigating not obviously and automatically be a recusable "conflict of interest"?

**Conclusion**

During this period of intense (deserved) scrutiny of various aspects of church governance and culture from many quarters, it has been common for commentators to think instinctively in relational terms. Even when focus of attention has been on specific, named individuals, discussion has often looked beyond personal misdeeds and culpability to *also* consider the extent to which individuals are enmeshed within, and thus both influence and influenced by, wider relationships and networks – and, therefore, also to the cultures, practices, and norms which they co-produce and co-maintain. Various examples of this from the reporting of the McCarrick case have been given above. Here let us add the earlier and more general observation of two Catholic legal scholars: "Bishops in the United States are part of a particular ecclesiastical culture that has its own influences, attitudes, and beliefs. This culture significantly influenced their decisions regarding clergy sexual abuse" (O'Reilly and Chalmers, 2014: 215). If this is so – and we firmly believe that it is – then tools which might help scholars, church officials, or indeed criminal investigators better understand the "social architecture" undergirding this culture are surely worth serious consideration.

In fact, our basic contention goes much deeper than this. An understanding of Catholic episcopal governance structures as "network[s] of social relations and interactions between actors" (Crossley 2011: 1) is not *just* a critical lens with which to investigate church scandals. Rather, social networks are integral to how such "ecclesiastical capital" (*pace* Bourdieu) is organized and maintained at all times, whether functioning well or not. Saintly and heroic bishops are, that is, just as much a "product" of networks as are defrocked former bishops. True, cases of abuse and cover-up serve as particularly apposite case-studies – both due to their clear and present "public interest" nature (i.e., in exploring how they have arisen, and how future ones might be prevented), and to the level of investigative journalism that has gone into uncovering the full details (on which we have drawn, gratefully and liberally, in offering our own interpretations). But they remain case studies supporting our much more wide-ranging claim. That is, the significance of SNA as one valuable, and hitherto untried, tool for understanding and/or improving Church governance. (As we hope is implicit in every paragraph of this paper, we do *not* think that SNA alone – uninformed by the insights of other forms of theological or social-scientific inquiry – is a kind of methodological "silver bullet" for this subject, or indeed any others.)

Given all this, we wish to conclude by suggesting six specific areas where we think a relational, network-informed approach has the potential to yield interesting fruit, several of which we are already working on/towards.

1. While we have privileged the Ordinary-subordinate tie here, there are of course others which may also be significant (Pogorelc, *forthcoming*). Among those that are, at least in theory, quantifiable, are: (co-)consecrators at episcopal ordinations; which seminaries/universities and years studied at; service together on certain committees or Boards of Directors. The degree to which these types of network overlap or correlate with each other would be a very interesting topic for enquiry.

2. Extending Ordinary-subordinate (or other types of) ties backwards through time. To what extent are "kingmakers" among the previous generations of bishops still significant over current episcopal politics? Is it useful to think in terms of episcopal "family trees", or even dynasties? Does the influence of different bishops or dioceses wax and wane over time (perhaps with changes in the pope or nuncio)?

3. Given the specific policies made by some bishops' conferences (USCCB included) for affirmative action in promoting ethnic or linguistic minority candidates, for example in the form of national lists, to what extent do these policies actually work?

4. Thinking practically and positively, if episcopal networks currently do exhibit tendencies towards certain "network pathologies", how might these same methods aid in reforming them? Would a policy of appointing "qualified outsiders" (i.e., suitable bishops not already tied into regional clusters) to major dioceses help in mitigating the kinds of conflicts of interest suggested above, especially in helping to "clean up" scandal hit dioceses? (Based solely on our Ordinary-subordinate dataset, the appointment of Archbishop Wilton Gregory to Washington as Wuerl's successor looks to be a promising example of precisely this.)

5. Moving beyond national episcopal politics, what might be gained from applying SNA methods to studying the Roman curia (itself a major object of scrutiny under the current pontificate). Which are really the powerful dicasteries? Which bishops, from which countries, sit on which especially influential Congregations (especially. the Congregation for Bishops)?

6. What light could SNA shed on certain historically important moments of episcopal politicking (cf. Wilde 2007; Pentin 2015; and O'Connell 2019 respectively)?

**Endnotes**

[1] I.e., (arch)bishops who are the head of a diocese or – in a small number of cases – who hold similar jurisdiction over a special grouping of Catholic clergy and laity outside of the normal Latin-rite diocesan structure. Examples of the latter include the "Personal Ordinariate of the Chair of St Peter", a special pastoral structure for certain ex-Protestant clergy and laity whose Ordinary is a bishop (though the Ordinaries of parallel Ordinariates in the UK and Australia are not: see note 4), and the various "Eparchies" of the Eastern Catholic Churches – that is, the jurisdictions of various autonomous Churches in communion with Rome, such as the Ukrainian Greek Catholic Church or the Syro-Malabar Catholic Church.

[2] Pogorelc's insightful chapter, a late-stage draft of which we gratefully received as we were putting the finishing touches to this paper, makes a parallel case to our own for the potential benefits of an SNA-informed approach to episcopal culture and governance, especially in light of the abuse crisis.

[3] In February 2019, the Vatican announced that, as the "conclusion of a penal process" in canon law, McCarrick had been found "guilty of […] solicitation in the Sacrament of Confession, and sins against the Sixth

Commandment with minors and with adults, with the aggravating factor of the abuse of power" (Congregation for the Doctrine of the Faith, 2019). He was accordingly laicized. Full details were not, however, made public. In this paper we have relied for such details on the reports of mainstream media sources (all cited).

[4] There are exceptions to this rule in the case of, for example, already-married ex-clergy from other denominations who convert to Catholicism and are accepted for Catholic ordination. But these are relatively few in number, and are, in any case, debarred from becoming bishops – and so sit outside of the kinds of network dynamics we're exploring here.

[5] 'Equivalents' here refers to other formal structures, existing in parallel to the territorial diocesan system, which possess their own Ordinary (whether a bishop or not). In England and Wales, these are the Bishopric of the Forces, two Eparchies of Eastern Catholic Churches (Syro-Malabar and Ukrainian), the Personal Ordinariate of Our Lady of Walsingham, and an Apostolic Prefecture responsible for the British Overseas Territories of the South Atlantic. Although not all of these are ordained as bishops, they are deemed "equivalent in [canon] law to a diocesan bishop" (Code of Canon Law, 381 §2). For ease of style and interpretation, in this paper all such "equivalents" will simply be included in the generic term "bishops", unless explicitly stated otherwise.

[6] Coadjutor status (roughly speaking, an auxiliary with right of episcopal succession) was originally coded separately as a "4". However, it was soon decided that this needlessly complicated matters (especially given the relative rarity of coadjutor appointments), and all 4s were recoded to 3s before analysis.

[7] Though beyond the scope of this paper, we are also exploring networked relationships within the Curia, using similar methods.

[8] "Historical" in the sense of relating to events that occurred often several years or decades ago. Their effects and implications are often, of course, very current.

[9] There are exceptions here, for example with central lists of suitable ethnic-minority candidates or the occasional cross-country posting of a priest with a national profile (as with Barron's appointment as auxiliary bishop of Los Angeles, from being a seminary Rector in Chicago), but these are just that: exceptions.